\documentclass[letterpaper, 10 pt, conference]{ieeeconf}  

\IEEEoverridecommandlockouts                              

\usepackage{xcolor}
\usepackage{amsmath}
\usepackage{amssymb}
\usepackage{mathtools}
\usepackage{dsfont}
\usepackage{comment}
\usepackage{algorithm} 
\usepackage{algpseudocode} 
\usepackage{caption}
\usepackage{graphicx}
\usepackage{cite}
\usepackage{hyperref}

\newtheorem{assumption}{Assumption}

\newtheorem{remark}{Remark}

\title{\LARGE \bf Data-Driven Modeling and Predictive Control of Chronic Diseases: An Ulcerative Colitis Application

}

\author{Fionna B. Kopp, Ann-Kathrin Schalkamp, Shadera Azzam, Vivek Rudrapatna, Francesco Borrelli
\thanks{Fionna B. Kopp and Francesco Borrelli are with the Department of Mechanical Engineering, University of California, Berkeley.}
\thanks{Fionna B. Kopp, Ann-Kathrin Schalkamp, Shadera Azzam, and Vivek Rudrapatna are with the Bakar Computational Health Sciences Institute, University of California, San Francisco.}
\thanks{This material is based upon work supported by the National Science Foundation Graduate Research Fellowship Program under Grant No. DGE 2146752. Any opinions, findings, and conclusions or recommendations expressed in this material are those of the author(s) and do not necessarily reflect the views of the National Science Foundation.}
\thanks{Research reported in this publication was supported by the
National Library of Medicine of the National Institutes of Health
under award number R00LM014099, as well as the National Center
for Advancing Translational Sciences, National Institutes of Health,
through UCSF-CTSI Grant Number UL1 TR001872. Federal sources
contributed approximately 10\% of the support for this study, with the remaining effort supported by non-federal sources. The content is solely the responsibility of the authors and does not necessarily represent the official views of the National Institutes of Health. This study was also supported by the Kenneth Rainin Foundation, and the department of medicine at University of California, San Francisco.}
}

\begin{document}

\maketitle
\thispagestyle{empty}
\pagestyle{empty}

\begin{abstract}
The clinical management of chronic disease is naturally structured as a feedback control problem. Physicians observe the disease state at clinic visits and make decisions based on the patient's history and current state. We propose a framework that models the dynamics of chronic disease in response to therapy as an event-based controlled stochastic system, in which treatment decisions are made at discrete, irregularly spaced events while the disease evolves between events as a continuous-time Markov chain (CTMC) with state and input dependent transition rates. We demonstrate this framework in the setting of ulcerative colitis, where we identify the model from real-world clinical data in the electronic health record (EHR) of more than 3,000 patients with ulcerative colitis. Using the learned model, we formulate therapy selection as a model predictive control problem (MPC) that maximizes the expected time spent in inactive disease states, and we evaluate the resulting policy. In closed-loop simulation on 10 patients, the MPC achieves a cost at or below the observed clinical course in the majority of its realizations for 4 of the 10 patients, and its outcomes in all cases are comparable to observed care and within clinical guidelines.
 
\end{abstract}

\section{Introduction}\label{sec:Intro}
Chronic diseases are managed over years through a sequence of longitudinal clinical decisions. At each clinical encounter, a physician assesses the patient's current disease state, considers how the patient has responded to previous therapies, and decides the next step in therapy. This naturally fits the framework of feedback control where the patient is the plant, the therapy is the input, clinical measurements of disease state are the output and the physician acts as the controller. Between encounters, the disease evolves stochastically and continuously, but is only observed at intermittent encounters.
 
Ulcerative colitis, a chronic, relapsing-remitting inflammatory disease of the colon, managed with long-term maintenance therapy, is a representative example~\cite{AGAGuidelines}. Clinical guidelines recommend a treat-to-target strategy, in which therapy is adjusted until a defined disease target is reached and then maintained~\cite{STRIDEII}. This is explicitly a feedback strategy. While the set of available therapies has grown substantially, standard clinical decision-making is based on general guidelines and randomized controlled trials (RCTs)~\cite{AGAGuidelines}, which offer limited direction on how to sequence therapies over time. This is in part because RCTs that compare therapies head-to-head are limited and RCTs generally do not evaluate multi-step treatment sequences~\cite{AGAGuidelines, VARSITY}. Clinical response in ulcerative colitis is marked by significant heterogeneity across patients and depends on past treatment history, most notably prior anti-TNF exposure~\cite{GEMINII_analysis, UNIFI_analysis, VARSITY}.
 
The EHR offers an opportunity to learn treatment response dynamics from at a scale and diversity that trials cannot match. However, EHR data is not collected for the purpose of modeling and several challenges arise from its inherent structure. The disease state is recorded only at clinical encounters, so the exact transition times between disease states are unknown and the trajectories are interval-censored panel data~\cite{KalbfleischLawless}. Encounters are irregularly spaced, making models that assume a fixed sampling interval are not directly applicable. In addition, information about a single quantity is often fragmented across several sources. Medication exposure, for example, appears in structured medication orders, in medication administration records, and in unstructured clinical notes, none of which is complete on its own.
 
Model-based control and sequential decision-making have been applied in many medical settings, including closed-loop glucose regulation~\cite{DoyleAP}, drug dosing and anesthesia~\cite{MPCAnesthesia}, and Markov decision process and reinforcement learning approaches for treatment selection learned from clinical records~\cite{Schaefer2004, sepsis}. Separately, continuous-time multistate Markov models are well established in biostatistics for describing disease progression from panel data~\cite{KalbfleischLawless}. Data-driven decision-making approaches typically discretize time into fixed steps, which is mismatched to irregularly timed visits, while multistate models are typically used for description and inference rather than for control. 
 
We model chronic disease response to therapy as an event-based controlled stochastic system. Between events, a discrete disease state evolves as a CTMC whose transition rate matrix depends on a mode determined by the current therapy, the patient's treatment history, and the time since the therapy was initiated. This mode structure captures the clinically known distinction between an induction phase, during which response to a new therapy is expected, and a maintenance phase. Because the model is defined in continuous time, transition probabilities over any interval follow from the matrix exponential, so irregular visit spacing is handled exactly. We identify the model from a large single-center cohort of patients with ulcerative colitis. To construct reliable state-input trajectories, we reconcile medication orders, medication administration records, and medication mentions extracted from clinical notes with a large language model (LLM) into a single medication timeline per patient. We then use the identified model in an MPC that, at each clinical visit, selects a therapy sequence that maximizes the expected time spent in inactive disease over a prediction horizon, subject to clinically relevant constraints.


The primary contributions of this paper are:
\begin{enumerate}
    \item We propose a general framework to model the dynamics of chronic disease response to therapy as an event-based controlled stochastic system, in which treatment decisions are made at discrete, irregular events while the disease evolves as a continuous-time Markov process with state and input dependent dynamics.
    \item We learn the model from state-input trajectory data extracted from the EHR by estimating the state and input dependent transition rate matrices for the continuous-time Markov process using the expectation-maximization algorithm.
    \item Using the data-driven model, we formulate therapy selection as an event-based MPC problem.
    \item We demonstrate the framework in the context of clinical management of ulcerative colitis.
\end{enumerate}
 
 
The remainder of the paper is organized as follows. In Sec.~\ref{sec:PD}, we introduce the event-based stochastic system framework and describe the specifics for our application in ulcerative colitis. Section~\ref{sec:modelID} describes the EHR data and model identification procedure. In Sec.~\ref{sec:MPC}, we formulate therapy selection as an event-based MPC problem. Section~\ref{sec:results} presents the identified model and results in the context of the clinical management of ulcerative colitis.
 
 

\section{Problem Statement} \label{sec:PD}
We consider an event-based stochastic system with dynamics
\begin{equation} \label{eq:dyn}
    x_{k+1} \sim P(\cdot | x_k, u_k, \Delta t_k), \quad \Delta t_k = t_{k+1} - t_k
\end{equation}
where $x_k \in \mathbb{R}^{n_c} \times \{0,1\}^{n_d}$ is the state at event $k$, $u_k \in \mathcal{U}$ is the input at event $k$, where $\mathcal{U}$ is a finite set, and $\Delta t_k$ is the relative time between event $k+1$ and event $k$. The corresponding system output at event $k$ is
\begin{equation}\label{eq:output}
    y_k = g(x_k)
\end{equation}
where $y_k \in \mathbb{R}^{p_c} \times \{0, 1\}^{p_d}$.

We assume the system satisfies the Markov property. The transition kernel $P(\cdot | x_k, u_k, \Delta t_k)$ specifies the conditional distribution of the state at the next event given the current state and input and the relative time between the events. We consider the case where transition kernel $P$ is unknown and is learned from state-input trajectories (Section~\ref{sec:modelID}).

We use the form specified by Eqs.~\eqref{eq:dyn} and~\eqref{eq:output} to model the response of a chronic disease to therapy, where the state $x_k$ captures characteristics of the patient's disease at event $k$, the input $u_k$ is the therapy applied from $[t_k, t_{k+1})$, and the transition kernel $P(\cdot | x_k, u_k, \Delta t_k)$ describes the stochastic evolution of the disease state in response to therapy. This framework applies to disease-management settings characterized by decisions made at discrete and irregularly-spaced events with stochastic disease progression and intermittently observed outcomes. In this paper, we apply this framework specifically to model response to therapy in the clinical management of ulcerative colitis. In Sec.~\ref{subsec:model_state_inputs} and Sec.~\ref{subsec:markov_model} we specify the state, input, output, and model for this application. In Section~\ref{sec:modelID} we describe a data-driven approach to learn the unknown transition kernel $P(\cdot | x_k, u_k, \Delta t_k)$ from state-input trajectories across a cohort of patients extracted from the electronic health record (EHR). Using the learned model, we design a model predictive controller (MPC) that selects the sequence of inputs $u_k$ to drive the output $y_k$ toward positive health outcomes. We present the data-driven model and the resulting control policy in Sec.~\ref{sec:results}.


\subsection{Modeling Ulcerative Colitis: Events, States, and Inputs} \label{subsec:model_state_inputs}
The structure of the model we have introduced (Eqs.~\eqref{eq:dyn} and~\eqref{eq:output}) is a general framework compatible for modeling chronic disease response to therapy and for subsequent use in a model-based controller to learn control (treatment) policies. The framework can be specialized with clinical and domain expertise for application. In this section, we describe our modeling choices for the application of the clinical management of ulcerative colitis, a chronic autoimmune disease affecting the gastrointestinal system.

In the context of managing ulcerative colitis, events $k$ are gastroenterology clinic visits at time $t_k$ where therapeutic decisions $u_k$ can be made including starting or stopping medications or continuing on the same regimen.

The input at event $k$ denoted by $u_k \in \mathcal{U}$ is the therapy regimen where $\mathcal{U}$ is the finite set of the modeled therapy regimens.

For the application of modeling response to therapy in ulcerative colitis, we define the state at event $k$ as
\begin{equation}
    x_k = \begin{bmatrix}
        \tau_k, 
        a_k,
        s_k,
        z_k,
        h_k
    \end{bmatrix}^\top
\end{equation}
where $\tau_k \in \mathbb{R}_{\geq 0}$ is the time since the current therapy regimen was initiated,
\begin{equation}
    \tau_{k} = \begin{cases}
        0, & u_k \neq u_{k-1} \\
        \tau_{k-1} + \Delta t_{k-1}, & u_k = u_{k-1},
    \end{cases}
\end{equation}
where we initialize $\tau_0 = \tau_{\text{init}}$ (Remark~\ref{rmk:init}). Activity $a_k \in \{0,1\}$ encodes the current disease as inactive or active and is derived from clinical measures recorded in the EHR. Severity $s_k \in \{1, 2, 3\}$ denotes the historical disease severity, which is monotonically non-decreasing and indicates the maximal severity (1 - mild / 2 - moderate / 3 - severe) to date as a proxy for difficult-to-treat or refractory disease. Severity is also derived from clinical measures in the EHR.
The medication history $h_k \in \{0, 1\}^{n_{\text{med}}}$ encodes prior exposure to the $n_{\text{med}}$ maintenance medications 
and $z_k \in \{0, 1\}$ indicates whether the patient has been exposed to anti-TNF therapy 
prior to the current therapy regimen.

To define $h_k$ and $z_k$, let $m_k \in \{0, 1\}^{n_{\text{med}}}$ encode the maintenance medications at event $k$ where $m_{k,i} = 1$ indicates that the medication represented by the $i$-th element is active. The encoding $m_k$ is constructed from information recorded in the EHR,
and the medication specific encoding $m_k$ contains more information than therapy regimen $u_k$ where the mapping from $m_k$ to $u_k$ is defined as $u_k = \phi_u(m_k)$.
We then define prior medication history $h_k$ as follows
\begin{equation}
    h_k = h_{k-1} \lor (m_{k-1} \land \neg m_k)
\end{equation}
where $\lor, \land,$ and $\neg$ denote the element-wise logical OR, AND, and NOT operations respectively, and we initialize with $h_0 = h_{\text{init}}$ (Remark~\ref{rmk:init}).

Since anti-TNF agents are a subset of the medications encoded in medication history, $z_k$ is a projection of $h_k$,
\begin{equation}
    z_k = \bigvee_{i \in \mathcal{Z}} h_{k,i}
\end{equation}
where $h_{k,i}$ is the $i$-th element of $h_k$ and $\mathcal{Z}$ is the set of indices corresponding to anti-TNF agents.

\begin{remark}\label{rmk:init}
    The time since starting the current therapy regimen is initialized as $\tau_0 = \tau_{\text{init}}$ and the medication history is initialized as  $h_0 = h_{\text{init}}$ where $h_{\text{init}}$ encodes medication exposure prior to event $0$. In this paper we use a single-center patient cohort, so $\tau_{\text{init}}$ and $h_{\text{init}}$ represent time on the active therapy regimen at the index visit and medication history prior to the index visit at the study center respectively.
    Generally the information needed to determine $\tau_{\text{init}}$ and $h_{\text{init}}$ is documented in clinical notes, and we extract it into usable data with a large language model (LLM).
\end{remark}

\begin{remark}
    Although $z_k$ is a function of $h_k$, we explicitly include both $z_k$ and $h_k$ in the state $x_k$. It is well-established that prior anti-TNF exposure reduces response rates to subsequent therapies, hence $z_k$ is used to determine the relevant transition probabilities in the Markov model~\cite{AGAGuidelines, GEMINII_analysis, VARSITY, UNIFI_analysis}. The full medication history $h_k$ is used in the control architecture (Section~\ref{sec:MPC}) to construct state-dependent input constraints, reflecting the common clinical practice of not returning to a previous therapy unless reused in combination therapy~\cite{AGAGuidelines}.
\end{remark}

The output of our system at event $k$ is a function of the state, defined as $y_k = g(x_k) = \begin{bmatrix}
    0 & 1 & 0 & 0 & 0 
\end{bmatrix} x_k = a_k$.

Next, we introduce the Markov model structure that we assume for system~\eqref{eq:dyn}.

\subsection{Markov Model} \label{subsec:markov_model}


\begin{assumption} [Markov property] \label{assm:markov}
    For the chosen state and input parameterization, the state at the next event depends only on the current state, input, and the time between events
    \begin{equation*}
        \begin{aligned}
            & P(x_{k+1} | x_k, u_k, \Delta t_k, x_{k-1}, u_{k-1}, \dots, x_0, u_0) \\
            & \qquad = P(x_{k+1} | x_k, u_k, \Delta t_k).
        \end{aligned}
    \end{equation*}
\end{assumption}

We partition the state $x_k = [\xi_k^\top, \eta_k^\top]^\top$ into the stochastically evolving component $\xi_k$ and the deterministically evolving component $\eta_k$. In the context of modeling therapy response in ulcerative colitis, $\xi_k = [a_k, s_k]^\top \in \Xi$, where $\Xi$ is the finite and discrete activity-severity space, and $\eta_k = [\tau_k, z_k^\top, h_k^\top]^\top$. The transition kernel $P$ from~\eqref{eq:dyn} factors into the deterministic part: $\eta_{k+1} = f_\eta(\cdot)$ (Sec.~\ref{subsec:model_state_inputs}), while the stochastic component evolves according to the dynamics of the CTMC. The propagation of $\xi_k$ over the interval $\Delta t_k$ is governed by a transition rate matrix $Q_{c_k} \in \mathbb{R}^{|\Xi| \times |\Xi|}$, where mode $c_k = \phi_c(x_k, u_k)$ is a deterministic function of the current state and input. The transition probabilities over the interval $\Delta t_k$ are then given by
\begin{equation}\label{eq:CTMC}
    P_\xi(\xi_{k+1} | x_k, u_k, \Delta t_k) = [\exp(Q_{c_k} \Delta t_k)]_{\xi_k, \xi_{k+1}}
\end{equation}
where $[\cdot]_{\xi_k, \xi_{k+1}}$ denotes the entry of the matrix indexed by the current activity-severity state $\xi_k$ and the next activity-severity state $\xi_{k+1}$. That is, Equation~\ref{eq:CTMC} extracts the probability of transitioning from $\xi_k$ to $\xi_{k+1}$ over interval $\Delta t_k$ in mode $c_k = \phi_c(x_k, u_k)$.

\begin{assumption} \label{assm:constant_rates}
    Within a given mode $c_k$, the transition rate matrix $Q_{c_k}$ is constant.
\end{assumption}

Since the severity component of the state is defined to be the maximal disease severity (1 - mild / 2 - moderate / 3 - severe) to date, the severity is monotonically non-decreasing for a given patient over time. This imposes structure on the transition rate matrices. Specifically, let $q_{ij}$ denote the elements of a transition rate matrix $Q$, where $q_{ij}$ is the transition rate from state $i$ to state $j$. Then for any state $j$ with corresponding severity strictly less than the severity corresponding to state $i$, $q_{ij}=0$ by definition.

Next, we describe the state and input dependent mode $c_k = \phi_c(x_k, u_k)$ in more detail. In the context of ulcerative colitis, the mode $c_k$ is defined by the time since the current therapy regimen was initiated, prior anti-TNF agent exposure, and the current input (therapy regimen): $c_k = (\tau_k, z_k, u_k)$. The named three components correspond to the axes along which the dynamics of clinical response are heterogeneous. The dependence on time since the current therapy regimen was initiated $\tau_k$ represents that the clinical response is not constant over time.
In therapeutic management of ulcerative colitis, medications are considered to have induction phases (typically on the order of 6-12 weeks) followed by maintenance. If a positive clinical response is going to occur as a result of therapy, the clinical response is generally observed during the induction period and then the goal of continued treatment is to durably maintain that response through long-term maintenance~\cite{AGAGuidelines, STRIDEII}. To reflect these temporally dependent dynamics, we partition the mode along the time since the current therapy was initiated into an \textit{induction} and \textit{maintenance} phase. The mode-dependence on prior anti-TNF exposure $z_k$ and current input $u_k$ reflects that prior anti-TNF exposure alters response rates and that therapeutic regimens have different efficacies~\cite{AGAGuidelines, GEMINII_analysis, UNIFI_analysis, VARSITY}.

\begin{remark}
    The mode $c_k$ and its dependence on the state and input through $\phi_c(\cdot, \cdot)$ are specified from domain expertise and clinical knowledge. We specified our choices for ulcerative colitis, but the framework is adaptable to other disease and therapy models with state and input definitions relevant to the clinical setting.
\end{remark}

We learn a separate transition rate matrix $Q_{c_k}$ for each mode $c$ from the transitions observed under that mode in the observational data set pulled from the EHR. In Section~\ref{sec:modelID} we describe the data set and how we learn the transition rate matrices.

We consider the Markov model as a controlled process where input $u_k$ is chosen at each event $k$. By choosing input $u_k$ the dynamic mode is determined through $c_k = \phi_c(x_k, u_k)$ which governs the stochastic propagation of the activity-severity component $\xi_k$ over the interval $\Delta t_k$. The control objective is to select the input sequence $u_k, u_{k+1}, \dots$ to maximize the probability that the activity-severity state $\xi_k$ reaches and remains in the inactive disease subset $\Xi_{\text{inactive}} \subset \Xi$. We formulate this as a model predictive control (MPC) problem in Section~\ref{sec:MPC}.

\section{Model Identification} \label{sec:modelID}
We learn the transition rate matrix $Q_{c_k}$ in~(\ref{eq:CTMC}) for each mode $c$ from observed data recorded in the EHR. In this section, we describe the information from the EHR that we use and how we apply the EM algorithm to learn the transition rate matrices.

\subsection{Electronic Health Record Data} \label{subsec:modelID_EHRdata}
The clinical history and activity of ulcerative colitis can be assessed from a number of markers recorded in the EHR.
Current and prior medication use is ascertained through the (a) record of medication orders, (b) administration events, and (c) extracting mentions of medications from clinic notes via an LLM~\cite{Vivek_LLM}. By reconciling these three sources a cohesive medication timeline for each patient is constructed.

At each event $k$, the state $x_k$ is constructed from the EHR: the activity severity component $\xi_k = [a_k, s_k]^\top$ is derived from clinical observations
and is fully determined at every event.
The remaining components of the state, $\tau_k, z_k,$ and $h_k$ along with the input $u_k$ are derived from the medication encoding $m_k$, where $m_k$ is constructed by evaluating the reconciled medication timeline at time $t_k$.
By evaluating the state $x_k$ and input $u_k$ at each event time $t_k$, we obtain a sequence of states and inputs for each patient $m \in \{1,\dots,M\}$
\begin{equation} \label{eq:state_input_trajs}
    \begin{aligned}
        \mathbf{x}^m & = [x_0^m, x_1^m \dots, x_{k}^m, \dots, x_{n_m}^m] \\
        \mathbf{u}^m & = [u_0^m, u_1^m \dots, u_{k}^m, \dots, u_{n_m-1}^m].
    \end{aligned}
\end{equation}
Due to the discrete and irregularly-spaced state observations, the observed trajectories are interval-censored observations of the underlying continuous time Markov process. We estimate the dynamics of the CTMC from the interval-censored trajectories~\eqref{eq:state_input_trajs} from $M$ patients; we describe this process in Sec.~\ref{}.

\begin{remark}
    For model identification, events $k$ at times $t_k$ are clinical encounters and medication starts or stops. This set of events consists of when state measurements occurred and when inputs changed. For the control framework (Section~\ref{sec:MPC}), events are the clinical encounters alone, allowing input changes to be made at encounters only.
\end{remark}

\subsection{Induction Period Identification}
The transition rate matrix $Q_{c_k}$ for the CTMC model is piecewise-constant along the time since initiating therapy axis $\tau_k$ with two-regimes: an induction phase and a maintenance phase. The mode $c_k$ is switched along three axes, $\tau_k$ being one of them. By modeling the induction period dynamics, we determine where to set the mode-switch along the temporal axis. Although the CTMC model represents all disease state dynamics, we define the temporal switch based on induction response as it is the expected dominant source of time-varying dynamics.

Medications used to treat ulcerative colitis are considered to have an \textit{induction period}, often on the order of 6-12 weeks, particularly for advanced therapies~\cite{STRIDEII}. Clinical trials often reflect this in their design by defining one of the primary endpoints at the end of the induction period. During this period the response from an initial active disease state to an inactive disease state is high. As time since initiating therapy $\tau_k$ increases, the patients remaining in an active disease state are increasingly likely to be non-responders to the medication, so the transition rate falls in the observed active disease state population.

We estimate these response dynamics as a first-order system where the transition rate decays toward a plateau as $\tau_k$ increases. Let $q(\tau)$ denote the active-to-inactive transition rate as a function of time since initiating therapy. We model the decaying transition rate as
\begin{equation} \label{eq:decaying_hazard}
    q(\tau) = q_{\infty} + (q_0 - q_{\infty})\exp \left (\frac{-\tau}{\tau_{\text{ind.}}} \right )
\end{equation}
where $q_0$ is the initial transition rate, $q_{\infty}$ is the maintenance conversion rate, and $\tau_{\text{ind.}}$ is the time constant corresponding to the induction period.

Induction response rates are specific to the active-to-inactive transition, so we restrict the data used to fit Eq.~\eqref{eq:decaying_hazard} to state-input trajectories following $\tau_k = 0$ (new therapy initiation) when $a_k = 1$ (active disease state at new therapy initiation). We consider the sequence of intervals following therapy initiation in an active disease state up to and including the first active-to-inactive transition. Subsequent intervals are not included as the induction response has already been observed, and subsequent dynamics including loss of response act at longer time scales. We refer to the set of considered intervals as the \textit{induction intervals}. 

We pool the fit across all therapies considered, reflecting a shared pharmacodynamic first-order response. Therapy dependent response rates are captured separately through mode $c_k$, which also switches along the input $u_k$ (therapy) axis.

EHR data is naturally panel-data, meaning that the state is observed only at discrete events $k$ and the exact time of an active-to-inactive state transition is unknown~\cite{KalbfleischLawless}. To fit Eq.~\eqref{eq:decaying_hazard}, we write the likelihood function over the induction intervals by calculating the probability of each observed interval outcome, conditioned on its initial state. The integral of the instantaneous rate of transition $q(\tau)$ is the cumulative hazard: 
\begin{equation}
    \begin{aligned}
        H(\tau) & = \int_0^\tau q(t) dt \\
        & = q_{\infty}\tau + (q_0 - q_{\infty}) \tau_{\text{ind.}}\left(1-\exp \left (\frac{-\tau}{\tau_{\text{ind.}}}
        \right ) \right ).
    \end{aligned}
\end{equation}
This admits survival function $S(\tau) = \exp(-H(\tau))$ which gives the probability of remaining in an active disease state through therapy time $\tau$ given starting in an initial active disease state at $\tau = 0$~\cite{StatsFailureTimeTextbook}. For an observed interval from event $k$ with therapy time $\tau_k$ and activity $a_k = 1$, to event $k+1$ with therapy time $\tau_{k+1}$, the probability of the disease state remaining active at $\tau_{k+1}$, conditional on $a_k = 1$ is
\begin{equation}
   P (a_{k+1} = 1 \mid a_{k} = 1, \tau_k, \tau_{k+1}) = \frac{S(\tau_{k+1})}{S(\tau_k)},
\end{equation}
and the probability of having transitioned to an inactive disease state at $\tau_{k+1}$ is then $P (a_{k+1} = 0 \mid a_{k} = 1) = 1 - \frac{S(\tau_{k+1})}{S(\tau_k)}$.

Let $i$ denote the interval index and let $\delta_i \in \{0, 1\}$ indicate interval $i$ is a transition from active to inactive when $\delta_i = 1$, otherwise $\delta_i = 0$ for remaining active. Under the assumption that intervals are independent, then log-likelihood over the induction intervals is
\begin{equation} \label{eq:induction_loglik}
\begin{aligned}
    \ell(q_0, q_\infty, \tau_{\text{ind.}}) = \sum_i \left [ \delta_i \log \left ( 1 - \frac{S(\tau_{i+1})}{S(\tau_{i})} \right ) \right. \\
    \left . + (1-\delta_i) \log \left ( \frac{S(\tau_{i+1})}{S(\tau_{i})} \right ) \right ]
\end{aligned}
\end{equation}
which we maximize over $(q_0, q_\infty, \tau_{\text{ind.}})$~\cite{StatsFailureTimeTextbook}.

\begin{assumption}
    Over the induction intervals, we assume that intervals with active endpoints $a_k = 1$ and $a_{k+1}=1$ do not contain unobserved active-to-inactive-to-active transitions. That is, we assume the disease remained in an active state throughout. This is reasonable for the set of induction intervals, which are early in therapy time where clinic follow-up tends to be more frequent, making interval durations shorter.
\end{assumption}

We fit the first-order decaying transition rate of induction dynamics~\eqref{eq:decaying_hazard} from $13{,}661$ induction intervals from $2{,}109$ patients pooled across $6$ modeled therapies.
The identified parameters are in Table~\ref{tab:induction_dyn}. The initial transition rate $q_0$ is about 2.7 times the steady state rate $q_\infty$ with a time constant estimate of $\tau_{\text{ind.}} = 48$ days, consistent with reported 6-12 week induction periods.

\begin{table}[t!]
\centering
\caption{First-order Induction Dynamics Fit (Patient-Level Bootstrap, 1000 Replicates)}
\label{tab:induction_dyn}
\begin{tabular}{lcc}
Parameter & Estimate & 95\% CI  \\
\hline
$q_0$ [yr$^{-1}$]             & 1.02 & [0.76, 2.12] \\
$q_\infty$ [yr$^{-1}$]        & 0.38 & [0.32, 0.43]  \\
$\tau_{\text{ind.}}$ [days] & 47.75 & [12.60, 118.54] \\
\hline
\end{tabular}
\end{table}

To confirm that the first-order induction model is warranted over a constant transition rate model, we perform a likelihood-ratio test against the constant-rate model, which is nested in the decaying rate model~\eqref{eq:decaying_hazard} by setting $q_0 = q_\infty$. The test statistic is 
\begin{equation}
    \lambda_{\text{LR}} = 2(\ell_{\text{decay}} - \ell_{\text{constant}}) = 59.2,
\end{equation}
indicating a preference to the first-order decaying-rate model over a constant rate model.

While the decaying-rate model is preferred, the confidence interval on $\tau_{\text{ind.}}$ is wide and on the order of the observation cadence, making estimating a continuously varying rate difficult at this timescale. We use the results of the first-order model fitting to locate the induction period, and approximate the dynamics with a piecewise constant CTMC~\eqref{eq:CTMC} switched at $\tau_{\text{switch}} = 2$ months.

\subsection{EM Algorithm}
We estimate the mode-dependent transition rate matrices $Q_c$ for all considered modes from state-input trajectories~\eqref{eq:state_input_trajs} observed in the EHR. The deterministic components of the state: $\tau_k, z_k,$ and $h_k$ evolve as established in Sec.~\ref{subsec:model_state_inputs}. Mode $c_k$ is a function of the deterministic components of the state as well as input $u_k$. The stochastically evolving component of the state $\xi_k$ has dynamics governed by the mode $c_k$ and its corresponding transition rate matrix~\eqref{eq:CTMC}.

We use the expectation-maximization (EM) algorithm to estimate the transition rate matrices from panel observations~\eqref{eq:state_input_trajs} where the stochastic component of the state $\xi_k$ (activity-severity in the context of ulcerative colitis modeling) is only measured at discrete events and the full trajectory between events is unobserved~\cite{EM_BladtSorensen}. Given observed endpoints of an interval $\xi_k$ and $\xi_{k+1}$, the probability of transition to $\xi_{k+1}$ from $\xi_k$ over the interval period $\Delta t_k$ in mode $c_k(x_k, u_k)$ is given by~\eqref{eq:CTMC}. Under Assumptions~\ref{assm:markov} and~\ref{assm:constant_rates}, and assuming that the $M$ patients in the cohort are independent, the log-likelihood is
\begin{equation} \label{eq:loglik}
    \ell(\{Q_c\}) = \sum_{m=1}^{M} \sum_{k=0}^{n_m-1} \log \left ( [\exp (Q_{c_k} \Delta t_k^m )]_{\xi_k, \xi_{k+1}} \right )
\end{equation}
where $\{Q_c\}$ denotes the set of all transition rate matrices.

\begin{remark}
    Note that an interval can be governed by multiple modes if $\tau_k \leq \tau_{\text{switch}}$ and $\tau_{k+1} > \tau_{\text{switch}}$, but the $(u_k, z_k)$ pair corresponding to the modes is constant throughout the interval. This allows us to decompose the log-likelihood by the $(u_k, z_k)$ components of the mode and fit the induction and maintenance periods across $(u_k, z_k)$ jointly, where intervals spanning $\tau_{\text{switch}}$ inform both the induction and maintenance dynamics.
\end{remark}

We maximize the log-likelihood~\eqref{eq:loglik} using the EM algorithm. We describe the EM iteration for a single $(u_k, z_k)$, jointly fitting across $\tau_{\text{switch}}$, and the same process is applied to each pair $(u_k, z_k)$ present in the set of considered modes. For brevity, let $Q$ denote the transition rate matrix being estimated. The EM algorithm alternates between computing the expected sufficient statistics under the currently estimated $Q$ and updating $Q$ from the expected sufficient statistics until convergence~\cite{EM_BladtSorensen}. We describe the approach briefly here, but defer to~\cite{EM_BladtSorensen} for full detail.

\paragraph{Sufficient statistics} Were the full activity-severity trajectory observed, the transition rate matrix would follow from the number of transitions from state $j$ to state $l$ denoted $N_{jl}$ and the total amount of time spent in state $j$, denoted $R_j$. Then
\begin{equation}
\begin{aligned}
    q_{jl} &= \frac{N_{jl}}{R_j}, \quad j \neq l \\
    q_{jj} &= -\sum_{l \neq j} q_{jl}
\end{aligned}
\end{equation}
where $q_{jl}$ is the element of $Q$ in the $j$-th row and $l$-th column.

However, the full trajectory is not observed in~\eqref{eq:state_input_trajs}, making $N_{jl}$ and $R_j$ unknown. Instead, the E-step calculates the expected values of the sufficient statistics given the observed intervals and current estimate of $Q$.

\paragraph{E-step} Given an interval with observed endpoints $\xi_k$ and $\xi_{k+1}$ with duration $\Delta t_k$ and currently estimated transition rate matrix $Q$ the expected number of transitions from state $j$ to state $l$ during the interval and the expected time in each state $j$ conditioned on the observed endpoints are
\begin{equation} \label{eq:EstepN}
    \mathbb{E} \left [N_{jl} | \xi_k, \xi_{k+1}, \Delta t_k \right ] = \frac{q_{jl}I_{jl}}{[\exp(Q \Delta t_k)]_{\xi_k, \xi_{k+1}}}
\end{equation}
and
\begin{equation} \label{eq:EstepR}
    \mathbb{E} \left [R_{j} | \xi_k, \xi_{k+1}, \Delta t_k \right ] = \frac{I_{jj}}{[\exp(Q \Delta t_k)]_{\xi_k, \xi_{k+1}}}
\end{equation}
where
\begin{equation*}
    I_{jl} = \int_0^{\Delta t_k} \left [ \exp (Q \sigma) \right]_{\xi_k, j} \left [ \exp (Q (\Delta t_k - \sigma)) \right ]_{l, \xi_{k+1}} d\sigma.
\end{equation*}
We refer the reader to~\cite{EM_BladtSorensen} for derivation.

\paragraph{M-step} Let $\mathcal{I}$ denote the set of all observed intervals across patients occurring under a given $(u,z)$ pair for which $Q$ is being learned. Each interval $i \in \mathcal{I}$ has endpoints $\xi_i$ and $\xi_{i}'$ with duration $\Delta t_i$.

The M-step sums the sufficient statistics over the observed intervals $i \in \mathcal{I}$ and updates the transition rate matrix as follows
\begin{equation}
\begin{aligned}
    q_{jl} & = \frac{\sum_{i \in \mathcal{I}} \mathbb{E}\left[ N_{jl} | \xi_i, \xi_{i}', \Delta t_i \right]} {\sum_{i \in \mathcal{I}} \mathbb{E}\left[ R_{j} | \xi_i, \xi_{i}', \Delta t_i \right]}, \quad j \neq l \\
    q_{jj} & = -\sum_{l\neq j} q_{jl}.
\end{aligned}
\end{equation}

\paragraph{Empirical Initialization} We initialize the algorithm with the empirical transition counts and times in state, by treating the observed endpoints as if they were the direct transitions, ignoring any unobserved jumps. The E-step and M-step are then alternated until convergence.

For our application in modeling ulcerative colitis, the severity component of the state is defined to be monotonically non-decreasing. The observed data~\eqref{eq:state_input_trajs} reflects the monotonically non-decreasing severity by definition. The transitions from a higher to lower severity state are fixed at zero at initialization, so they contribute no expected transitions in the E-step and are not updated in the M-step, forcing these transition rates to remain fixed at zero through the EM iterations.

We present the fitted models and their bootstrapped confidence intervals for our application in Sec.~\ref{sec:results}.

\section{MPC} \label{sec:MPC}
Using the learned CTMC dynamics, we formulate therapy selection as a event-based MPC problem. In a clinical setting, therapy decisions are made at events $k$ (clinical visits) with the goal being to optimize health outcomes over a horizon. At the next event, a new observation of the disease state is made and therapy adjusted as needed to optimize outcomes. This naturally fits the framework of MPC. In this section, we formalize this where the MPC is subject to the learned dynamics~\ref{sec:modelID}.

At each event $k$, the controller selects optimal inputs (therapy) to optimize expected disease outcomes over a prediction horizon of $N$ events. The learned CTMC dynamics are used to predict the evolution of the stochastic activity-severity state component over the horizon.
At event $k$ the following finite-horizon problem is solved:
\begin{subequations} \label{eq:MPC}
    \begin{align}
        J_{k \rightarrow k+N}^{\text{MPC}} = & \min_{\mathbf{u}_k} \sum_{n=k}^{k+N-1} \mathbb{E} \left [ r (y_{n|k}) \right ] \Delta t_{n}  \nonumber \\
        & \quad  \quad +  \mathbb{E} \left [ r (y_{k+N|k}) \right ] \Delta t_{\text{term}} \nonumber \\
        & \quad \quad + r_{\text{switch}} \sum_{n=k}^{k+N-1} (1 - u_{n|k}^\top u_{n-1|k}) \label{subeq:MPC_cost} \\
        \text{s.t. } & \xi_{n+1|k} \sim P_{\xi} ( \cdot \mid x_{n|k}, u_{n|k}, \Delta t_n) \label{subeq:MPC_stochdyn}\\
        & \eta_{n+1|k} = f_\eta(\eta_{n|k}, u_{n|k}, u_{n+1|k}, \Delta t_n) \label{subeq:MPC_detdyn}\\
        & c_{n|k} = \phi_c(x_{n|k}, u_{n|k}), \label{subeq:MPC_mode} \\
        & \quad \quad \quad \forall n \in \{k, \dots, k+N-1 \} \nonumber\\
        & x_{n|k} = \left [\xi_{n|k}^\top, \eta_{n|k}^\top \right ]^\top \label{subeq:MPC_state}\\
        & y_{n|k} = g(x_{n|k}) \label{subeq:MPC_output}\\
        & u_{n|k} \in \mathcal{U}(x_{n|k}) \subseteq \mathcal{U} \label{subeq:MPC_input_constr}\\
        & \quad \quad \quad  \forall n \in \{k, \dots, k+N \} \nonumber\\
        & x_{k|k} = x_k, \label{subeq:MPC_IC_state}\\
        & u_{k-1|k} = u_{k-1} \label{subeq:MPC_IC_input}
    \end{align}
\end{subequations}
where $\mathbf{u}_k = \begin{bmatrix} u_{k|k}, \dots, u_{k+N-1|k} \end{bmatrix}$. The cost function~\eqref{subeq:MPC_cost} sums the expected cost of output over the horizon weighted by its interval duration, a terminal cost term on the expected cost of the output $N$ events into the future weighted by $\Delta t_{\text{term}}$, and a penalty on switching inputs weighted by constant $r_{\text{switch}}$. Equations~\eqref{subeq:MPC_stochdyn} and~\eqref{subeq:MPC_detdyn} propagate the state~\eqref{subeq:MPC_state} over the horizon. Equation~\eqref{subeq:MPC_mode} defines the state and input dependent mode, which drives the transition rate matrix governing the stochastic dynamics in~\eqref{subeq:MPC_stochdyn}. The output~\eqref{subeq:MPC_output} is a function of the state, which makes the output a distribution over which we take the expectation to formulate the cost. State-dependent input constraints are enforced through Eq.~\eqref{subeq:MPC_input_constr}. Initial conditions on the state and input are set in~\eqref{subeq:MPC_IC_state} and~\eqref{subeq:MPC_IC_input}.

Let $\mathbf{u}^*_k$ denote the optimal input sequence to~\eqref{eq:MPC} at event $k$. At event $k$, the first element of $\mathbf{u}^*_k$ is applied to system~\eqref{eq:dyn}: $u_k = u_{k|k}^*$. The FTOCP~\eqref{eq:MPC} is repeated at event $k+1$, with new state measurement $x_{k+1}$ in a receding horizon manner. Note that problem~\eqref{eq:MPC} is a mixed-integer due to the switched (mode-dependent) dynamics.

\begin{remark}
    In the observed data~\eqref{eq:state_input_trajs}, patients are frequently on multiple medications (concomitant medications) simultaneously, all encoded in $m_k$. For the purposes of defining the model and model identification from observed data, the medication history $h_k$ is computed from the encoding $m_k$ of the reconciled medication timeline (Sec.~\ref{subsec:model_state_inputs}). In the setting of the MPC, $h_k$ updates from the selected therapy sequence $u_k$ which does not encode concomitant medications. Hence, we can write the deterministic dynamics in the MPC~\eqref{subeq:MPC_detdyn} in terms of $u$.
\end{remark}

\section{Results} \label{sec:results}

We applied the proposed framework to the pharmacological management of ulcerative colitis. In the following subsections we describe the identification of the CTMC dynamics from EHR data (Sec.~\ref{subsec:results_modelID}) and we evaluate the MPC~\eqref{eq:MPC} in closed-loop on patients sampled from the cohort with the learned dynamics defining both the prediction model of the controller and the dynamics of the simulated system.

\subsection{Model Identification} \label{subsec:results_modelID}
We identify the model~\eqref{eq:CTMC} from a single-center cohort of adult patients with ulcerative colitis comprising of state-input trajectory data~\eqref{eq:state_input_trajs} from the EHR. The EHR was adopted in 2012, so all data is from 2012 and after.

For each mode $c$, we estimate the transition rate matrix $Q_c$ from the observed intervals governed by that mode. We considered all approved treatments for ulcerative colitis observed in the cohort and retained those with sufficient data to identify the CTMC model. Model identifiability was assessed by a patient-level bootstrap for each of the models with $500$ replicates. The models under siz therapies were sufficiently identified: infliximab, mesalamine, vedolizumab, ustekinumab, adalimumab, and tofacitinib. With the 2 temporal periods, $\tau \leq \tau_{\text{switch}}$ and $\tau > \tau_{\text{switch}}$, and 2 prior anti-TNF exposure states $z \in \{0, 1\}$, this results in $24$ identified transition rate matrices $Q_c \in \mathbb{R}^{6 \times 6}$ over the activity-severity space, each with patient-level bootstrap confidence intervals. The set of $24$ identified transition rate matrices were learned from trajectories of $3{,}318$ patients covering $33{,}706$ intervals. The bootstrap confidence intervals were used only to determine identifiability of the modes; while all results below use the nominal estimate in the MPC and simulated dynamics. We report more detail on the learned models and confidence intervals in~\cite{}.




\subsection{Closed-Loop MPC for Ulcerative Colitis} \label{subsec:results_MPC}
We implement the MPC~\eqref{eq:MPC} with the input set $\mathcal{U}$ as the set of six identified therapies and a prediction horizon of $N=4$ events. The output-dependent component of the cost function is $r(y_n) = -\mathds{1}[a_n = 0]$ where $\mathds{1}$ is the indicator function. We use a terminal weight $\Delta t_{\text{term}} = 3$~months and each change in therapy is penalized by $r_{\text{switch}} = 0.15$.

In the context of ulcerative colitis, the state-dependent input constraint $\mathcal{U}(x_n)$ in Eq.~\eqref{subeq:MPC_input_constr} encodes that a therapy in the medication history $h_n$ cannot be started again.

To evaluate the performance of our MPC in closed-loop, we applied the MPC in closed loop to $10$ patients from the cohort (in simulation) and compared the simulated outcomes under the MPC policy to their observed trajectories recorded in the EHR. Under this comparison, our baseline reflects the treatment that the patients received at the IBD center. Next we describe how we implemented this comparions.

We sampled $10$ patients from the cohort and an initial state for each patient was defined from their visits at which their current therpy was one of the six modeled therapies with known start date, and at least two follow-up visits occurred within the following 24 months over which they were on modeled therapies. Given these conditions, we construct a set of 10 initial states $x_0 = [\tau_0, a_0, s_0, z_0^\top, h_0^\top]^\top$ (one for each sampled patient). The initial state is fully determined by the information in the EHR. For each patient we also pull the corresponding sequence of follow-up intervals (at minimum two) for at most the next 24 months. Since events are irregularly space across intervals, the event spacing for each sampled patient is unique. For each patient we evaluate the following:
\begin{enumerate}
    \item \textit{Observed course:} We evaluate the cost~\eqref{subeq:MPC_cost} associated with the observed states and the input sequence actually given as recorded in the EHR with the event-spacing defined by the recorded follow-up visits. We denote the resulting cost $J_{\text{obs}}$.
    \item \textit{Closed-loop MPC:} At each event time pulled for the sequence of follow-ups (starting at the initial condition), we solve the MPC~\eqref{eq:MPC} and apply the first input, simulate the state forward to the next observed event time through the learned CTMC. At the next event time, the updated state measurement is drawn from the propagated distribution. For each patient we simulated their closed-loop MPC trajectory 10 times, giving a distribution of the closed-loop costs.
\end{enumerate}

\begin{figure}
    \centering
    \includegraphics[scale=0.37]{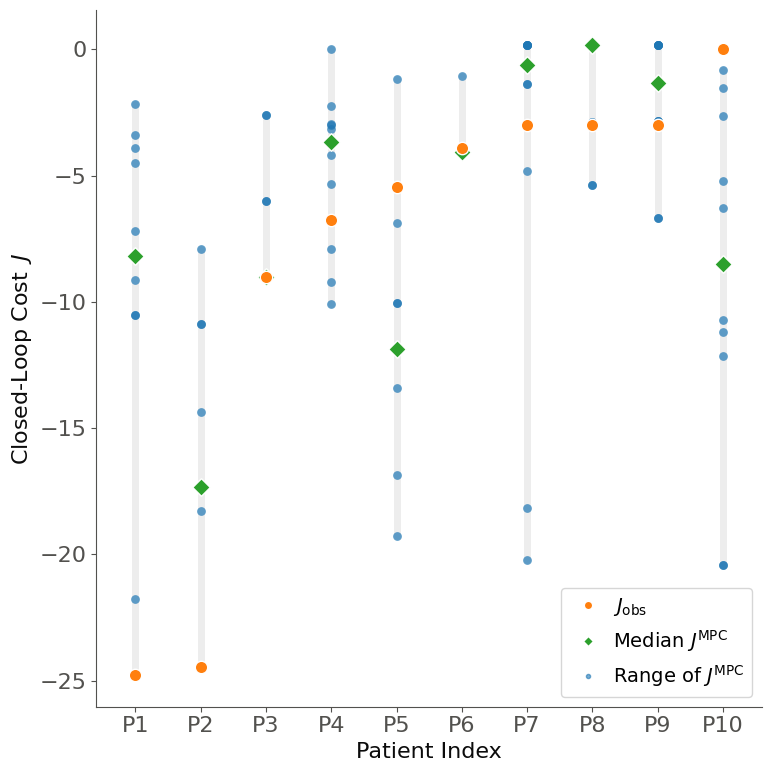}
    \caption{The observed closed loop cost $J_{\text{obs}}$ for each patient is denoted by an orange dot. The blue dots denote the 10 closed-loop MPC costs ($J^{\text{MPC}})$, and the light gray bar spans their range for each patient. The green diamonds denote the median closed-loop MPC cost for each patient.}
    \label{fig:CLcost}
    \vspace{-0.4cm}
\end{figure}

For each of the sampled patients, Fig.~\ref{fig:CLcost} compares the cost of the real observed course with the costs fo the 10 closed-loop MPC trajectories in simulation. The cost $J_{\text{obs}}$ of the observed course fell above or within the range of the closed-loop MPC costs for all 10 sampled patients. For four of the 10 patients, more than half of the simulated MPC closed-loop costs achieved a cost at or below the observed cost $J_{\text{obs}}$ (Fig.~\ref{fig:CLcost}). The closed loop MPC achieves comparable outcomes to the observed clinical course.

\begin{figure}
    \centering
    \includegraphics[scale=0.35]{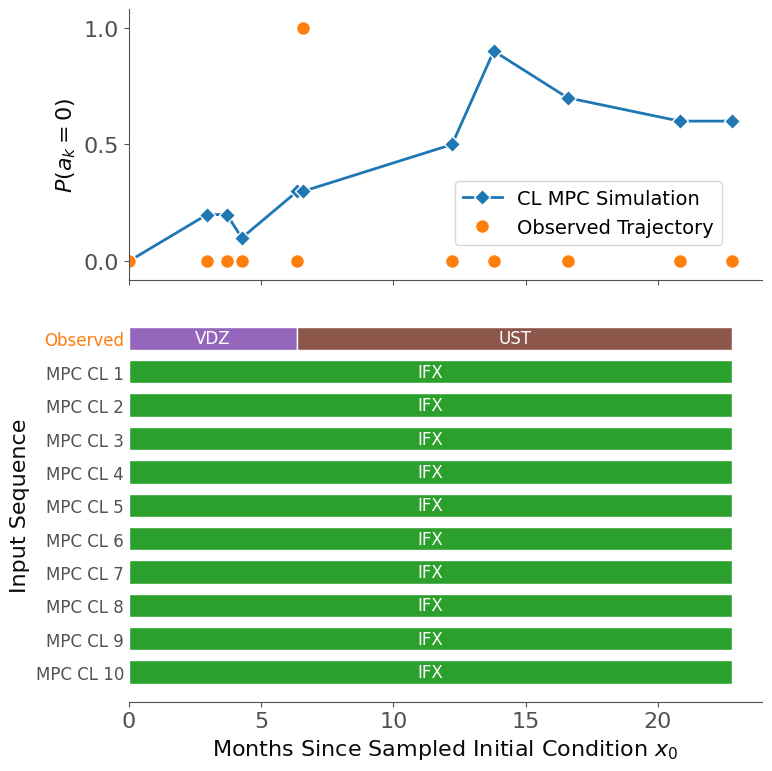}\hfill
    \includegraphics[scale=0.35]{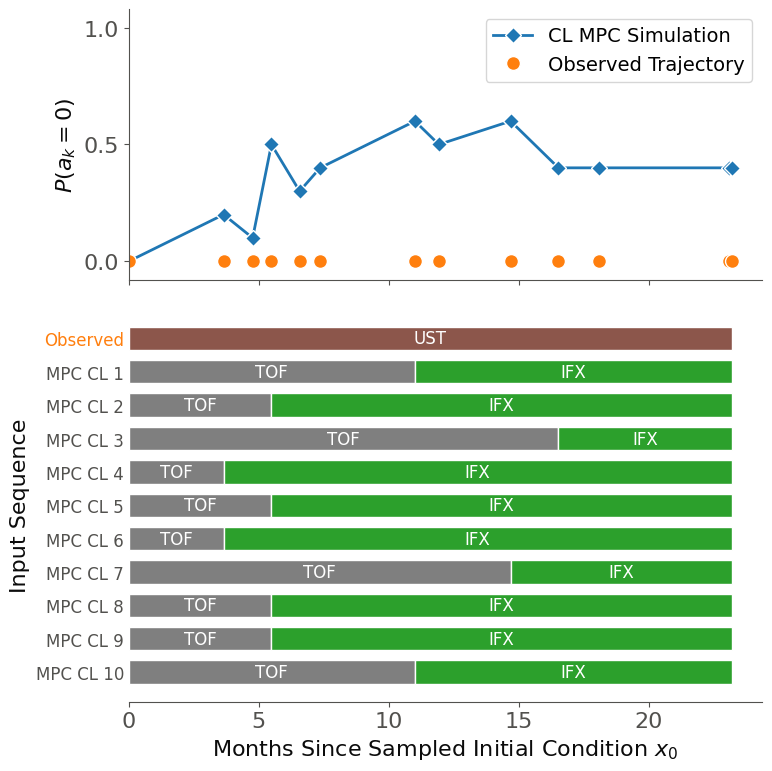}
    \caption{Observed course versus the closed-loop MPC simulations for two sampled patients. The top panel in each illustrates the remission at each event and the bottom panel shows the input sequence over time. The abbreviations in the figure are as follows: vedolizumab (VDZ), ustekinumab (UST), tofacitinib (TOF), and infliximab (IFX).}
    \label{fig:CLtrajs}
    \vspace{-0.4cm}
\end{figure}

In Fig.~\ref{fig:CLtrajs} we show trajectories of two of the sampled patients, including their observed state-input trajectory as well as the input sequence from the 10 closed-loop MPC simulations. In both cases, the patient was active at the initial state and the MPC recommended an alternative therapy. In both situations the recommended alternative therapy is within clinical guidelines; the MPC switched patients on advanced therapies to alternative advanced therapies rather than a medication from a lower drug class.
    
\section{Conclusion} \label{sec:conclusion}

We proposed a general framework that models the response of chronic disease to therapy as an event-based controlled stochastic system, with dynamics learned from state-input trajectories in the EHR. Using the learned model, we frame therapy selection as an event-based MPC problem while accounting for clinical constraints. We demonstrated the approach in the context of ulcerative colitis, where we identified the transition dynamics for six therapies from a single-center cohort of over $3{,}000$ patients. Evaluated in closed-loop, the MPC achieved outcomes comparable to the observed clinical courses and within clinical guidelines. Future work will use the bootstrap estimates of model uncertainty to construct a robust formulation and advance the development of data-driven clinical decision-support tools.

\bibliographystyle{ieeetr}
\bibliography{root}

\end{document}